# Electrical-Field Modulation of the Charge-Density-Wave Quantum Condensate in *h*-BN/NbS$_3$ Heterostructure Devices

Maedeh Taheri[1], Nicholas Sesing[2], Tina, T. Salguero[2], and Alexander A. Balandin[1,3]*

[1]Department of Materials Science and Engineering, University of California, Los Angeles, California 90095 USA

[2]Department of Chemistry, University of Georgia, Athens, Georgia 30602 USA

[3]California NanoSystems Institute, University of California, Los Angeles, California 90095 USA

### Abstract

**We report on the field-effect modulation of the charge-density-wave quantum condensate in the top-gated heterostructure devices implemented with quasi-one-dimensional NbS$_3$ nanowire channels and quasi-two-dimensional *h*-BN gate dielectric layers. The charge-density-wave phases and collective current in quasi-1D NbS$_3$ nanowires were verified *via* temperature dependence of the resistivity, non-linear current-voltage characteristics, and Shapiro steps that appeared in the device response under radio frequency excitation mixed with the DC bias. It was demonstrated that the electric field of the applied gate bias can reversibly modulate the collective current of the sliding charge-density-wave condensate. The collective current reduces with more positive bias suggesting a surface effect on the condensate mobility. The single particle current, at small source–drain biases, shows small-amplitude fluctuation behavior, attributed to the variations in the background potential due to the pinned or creeping charge-density-wave condensate. The knowledge of the electric-field effect on the charge density waves in quasi-1D NbS$_3$ nanowires is useful for potential electronic applications of such quantum materials.**

---

* Corresponding author (A.A.B.): balandin@seas.ucla.edu ; https://balandin-group.ucla.edu/

Low-dimensional van der Waals materials continue to attract growing attention owing to their unique properties associated with strongly correlated phenomena such as the charge-density-wave (CDW) quantum condensate phases and various topological and chiral effects [1-15]. The quasi-one-dimensional (1D) crystals of the transition metal trichalcogenides (TMT) with the formula MX$_3$ (M = Nb, Ta; X = S, Se), and specifically, NbSe$_3$ and TaS$_3$, are well-known from the studies of CDW effects conducted over the past few decades [16-24]. A member of this family, NbS$_3$, is another example of a material system with intriguing CDW properties, which reveals the quantum condensate phase at room temperature (RT) [25-27]. The RT or near-RT CDW phases, which can be controlled with the source–drain electrical field and, potentially, gate bias, are particularly interesting from both the physics and practical applications points of view. The CDW materials with quasi-1D crystal structures have been proposed for future radio frequency (RF) technologies, including detectors, mixers, and related communication devices [28-30].

The majority of the demonstrated prototype CDW devices are two-terminal. It was suggested that they have the potential for low-power and high-speed operation owing to the nature of CDW transitions [31]. The distortion of the crystal lattice during the CDW phase transitions is rather small and can be induced both by the local heating and electrical field [9, 32]. For these reasons, CDW devices are different from "conventional" resistive switches that rely on the Joule heating-induced transitions from polycrystalline to amorphous phases [33-35]. Achieving even a weak modulation of the CDW phases with the electrical gate would increase the application potential of CDW devices. The latter explains early attempts to create a device similar to the field-effect transistor (FET) with channels made of CDW materials such as NbSe$_3$, TaS$_3$, and K$_{0.3}$MoO$_3$ [23, 24, 36]. The reported gating of CDW current in materials with quasi-1D crystal structure was carried out in the large channel back-gated devices and with bulk CDW samples. For our study, we selected nanowires of quasi-1D crystalline NbS$_3$, and employed the top-gate device design with the gate dielectric comprised of boron nitride (*h*-BN). We investigated the effect produced by the gate bias on the device drain current in the pinned CDW condensate regime, *i.e.* for the source–drain electric field below the threshold field, $E_T$, and in sliding CDW condensate regime, *i.e.* after the CDW depinning and onset of the collective current. Our results demonstrate that the effect produced by the gate bias is strikingly different in these two cases.

Despite its interesting properties, NbS$_3$ material is less studied compared to NbSe$_3$, TaS$_3$, and some other TMT compounds. This contrast is rooted in the polymorphism of NbS$_3$, and the difficulties associated with the synthesis and isolation of single-phase samples [37-39]. Some polymorphs are metallic and reveal CDW phases, while others are semiconductors. The monoclinic polymorphs include NbS$_3$-II, a metal at higher temperature with three CDW phases at Peierls temperatures $T_{P0}$ = 460 K, $T_{P1}$ = 330 to 370 K, and $T_{P2}$ = 150 K [25, 27]. The CDW phase below $T_{P1}$ revealed high coherency under microwave irradiation and the potential of synchronization up to the frequency of 200 GHz [25, 29]. Here we focus on the NbS$_3$-II polymorph because it has well-defined transition temperatures and a CDW quantum condensate phase that extends above RT.

The NbS$_3$ crystals were synthesized using the chemical vapor transport (CVT) method with excess sulfur applied as a transport agent and a 973 K → 943 K gradient [37, 39]. Wirelike crystals were selected from the dark silver-grey crystals grown throughout the ampule. The scanning electron microscopy (SEM) image and the energy-dispersive X-ray spectroscopy (EDS) mapping of the NbS$_3$ crystal are shown in Figure 1 (a). The van der Waals bonding in the crystal facilitates mechanical and chemical exfoliation of the NbS$_3$ whiskers down to nanowires with diameters on the order of a few nanometers. In this study, the samples were mechanically exfoliated and micromanipulated on the SiO$_2$/Si substrate. The NbS$_3$ nanowires were capped with *h*-BN layers immediately after the exfoliation, using the all-dry transfer method. The *h*-BN cap layer provides channel protection from environmental exposure and serves as an effective gate dielectric [40, 41]. The source-drain contacts were fabricated using the electron-beam lithography, followed by dry-etch of the *h*-BN layer and Ti/Au metal deposition. The gate contact was patterned and metalized on top of the *h*-BN film. Figure 1 (b) shows a schematic of the device with the NbS$_3$ channel and the *h*-BN gate dielectric. The channel length, $L$, was on the order of a few micrometers while the cross-sectional dimensions, A, were below ~8×10$^{-3}$ μm$^2$. In Figure 1 (c), we present an SEM image of a representative top-gated *h*-BN/NbS$_3$ quasi-2D/1D heterostructure device, with Ti/Au (10 nm /90 nm) contacts. Additional materials and device characterization data are provided in Supplementary Figures S1 – S3.

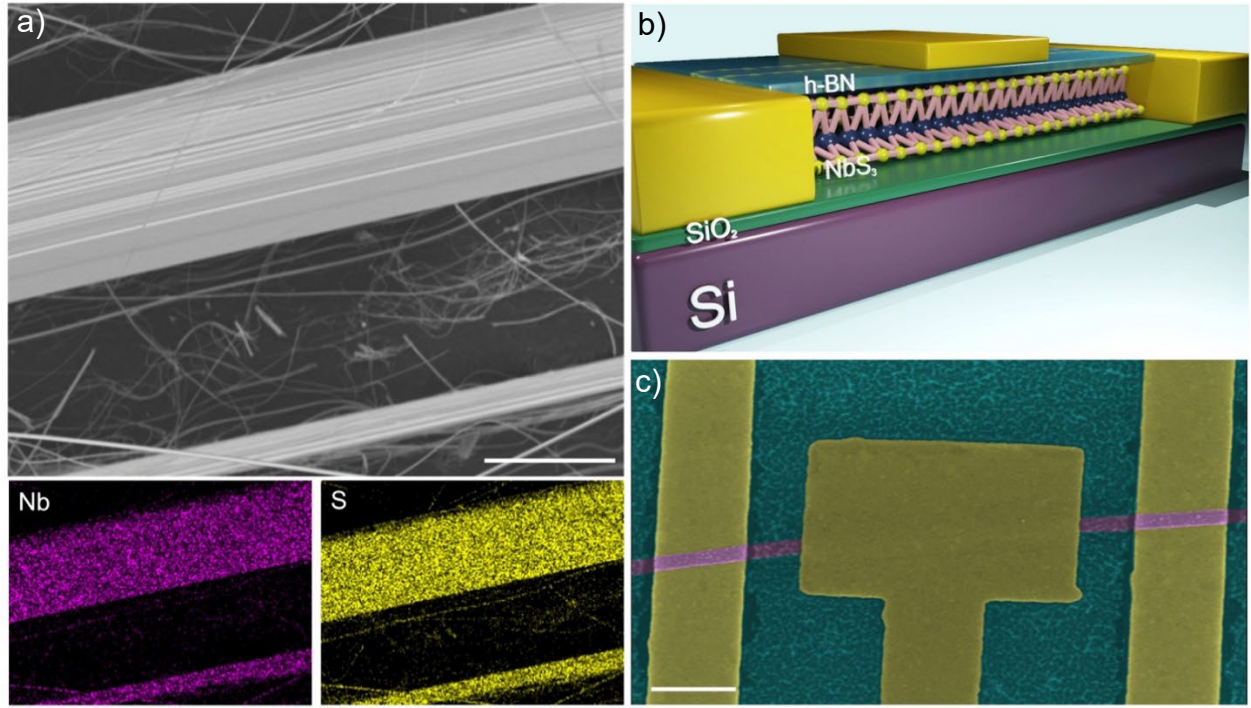

**Figure 1**: (a) SEM image and EDS elemental mapping of wirelike NbS$_3$ crystals. The scale bar is 25 μm. (b) Schematic of a top-gate NbS$_3$ device with *h*-BN gate dielectric layer. The nanowire of NbS$_3$ is depicted as a monoclinic atomic structure with S (yellow) and Nb (grey) atoms. (c) A representative SEM image of the *h*-BN/NbS$_3$ device. The pseudo colors are used for clarity. The scale bar is 1 μm.

The temperature dependence of NbS$_3$ channel resistance is plotted in a semilogarithmic scale in Figure 2 (a). The two anomalies in the resistance indicate the CDW phase transitions, which are shaded as $T_{P2} \approx 154 - 164$ K and $T_{P1} \approx 320 - 375$ K. The transition temperatures are in line with the type-II polymorph of NbS$_3$ and confirm the CDW quantum condensate states in the nanowires. Figure 2 (b) shows the current-voltage (*I-V*) characteristics for a few temperatures in the range between $T_{P1}$ and $T_{P2}$. The arrows indicate the threshold voltage, $V_T$, of the CDW depinning and on-set of the CDW sliding. The accurate on-set of the sliding was determined from the differential, *dV/dI*, characteristics, defined as the start of the resistance drop (see Figure 2 (c)). The nonlinearity in the *I-V* characteristic of the CDW materials with a quasi-1D crystal structure is explained within the "two-fluid" model of the contributions of individual single particle electrons and the sliding of the electron–phonon quantum condensate of the CDW phase [1, 16, 17, 19]. The depinning and sliding of CDW contribute to the total current and result in super-linear *I-V* curves for $V_D \geq V_T$.

The threshold field, $E_T$, of the CDW depinning, related to $V_T$, as $V_T = E_T \times L$, is a function of temperature. Figure 2 (d) shows the evolution of the threshold field as the temperature changes over two transition points, $T_{P1}$ and $T_{P2}$.

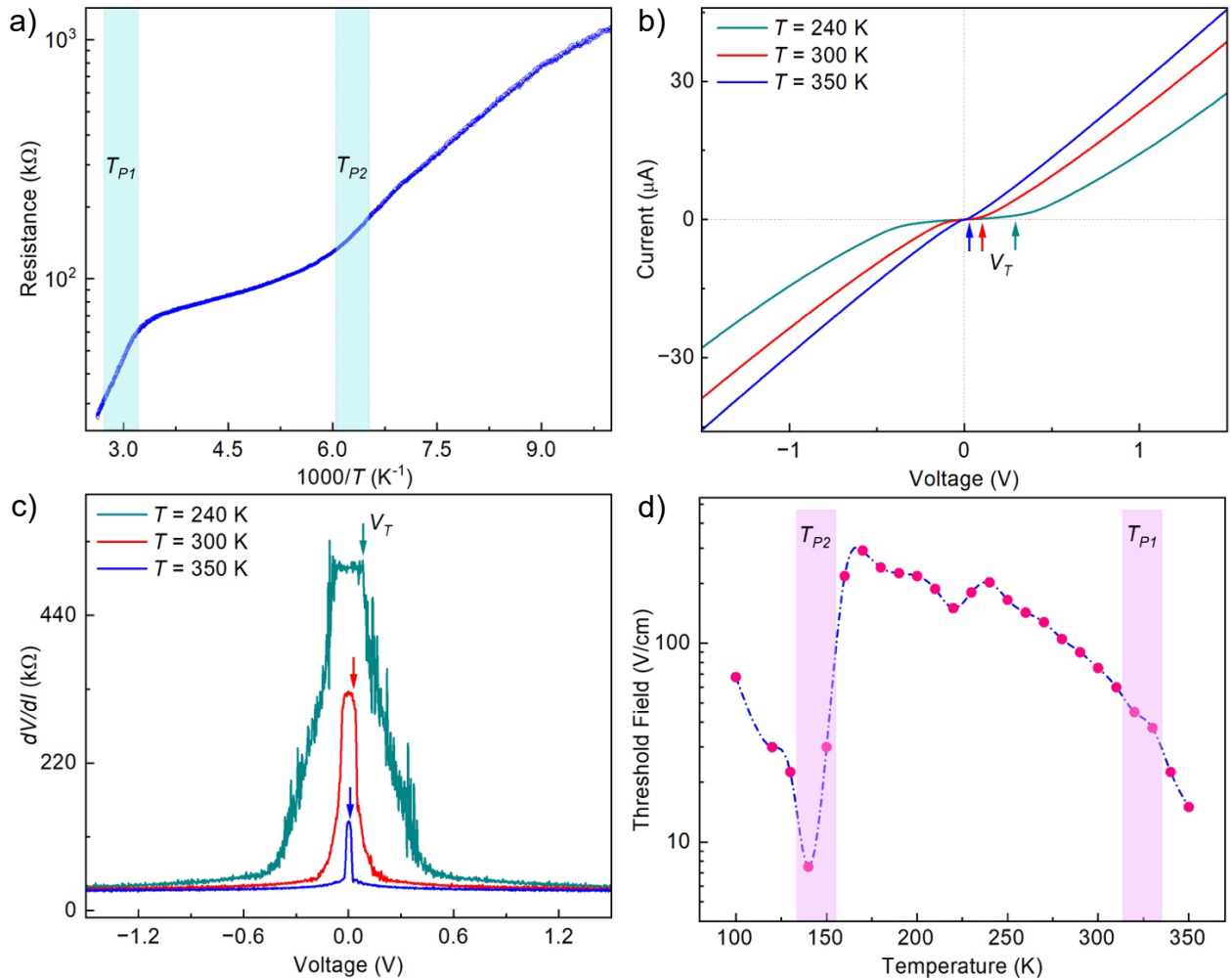

**Figure 2**: (a) Electrical resistance as a function of inverse temperature in log-linear scale. The anomalies in the resistance correspond to the CDW phase transitions: $T_{P1}$ = 320 K to 375 K, and $T_{P2}$ = 154 K – 164 K. (b) *I-V* characteristics of NbS$_3$ devices at three different temperatures $T$ = 240 K, $T$ = 300 K, $T$ = 350 K. The linear region is due to the single-particle conduction when the CDW quantum condensate is pinned and not contributing to the current. Beyond the threshold voltage, $V_T$, the CDW condensate starts sliding, leading to the non-linearity in the *I-V* curves. (c) The differential, $dV/dI$, characteristics *vs.* $V$, plotted for several temperatures. (d) Threshold field, $E_T$, as a function of temperature.

Typically, the temperature dependence of the resistance and non-linear *I-V*s with meaningful transition points are considered sufficient for proving the existence of the CDW phases. To provide additional evidence for CDW quantum condensate phases in our NbS$_3$ nanowire channels we also measured *I-V* characteristics with RF input. We consider this important for the study, which is focused on distinctions of the electrical modulation of CDW current before and after depinning. The approach for mixing the frequency of the sliding CDW with the external RF for monitoring CDW sliding has been implemented with different materials [16-19, 42-45]. The CDW sliding introduces collective current which the AC component termed the "narrow-band-noise" (NBN) [19]. When the NBN frequency, ω$_{NBN}$, is equal to or proportional to the frequency of an external microwave source, ω$_{ex}$, *i.e.*, $\omega_{NBN} = \omega_{ex}(p/q)$, where *p* and *q* are integers, the interference can occur. It reveals itself as steps in *I-V* characteristics, known as Shapiro steps, by an analogy with the phenomenon in superconductors. Once the RF synchronization occurs, the CDW velocity remains nearly constant within the step voltage range. The CDW fundamental frequency, *f$_0$*, is directly related to the CDW current, as $f_0 = I_{CDW}/MeN$, where *e* is the elementary charge, *N* is the number of atomic chains in the sample, and $M \approx 2$ is the number of electrons per CDW wavelength [25].

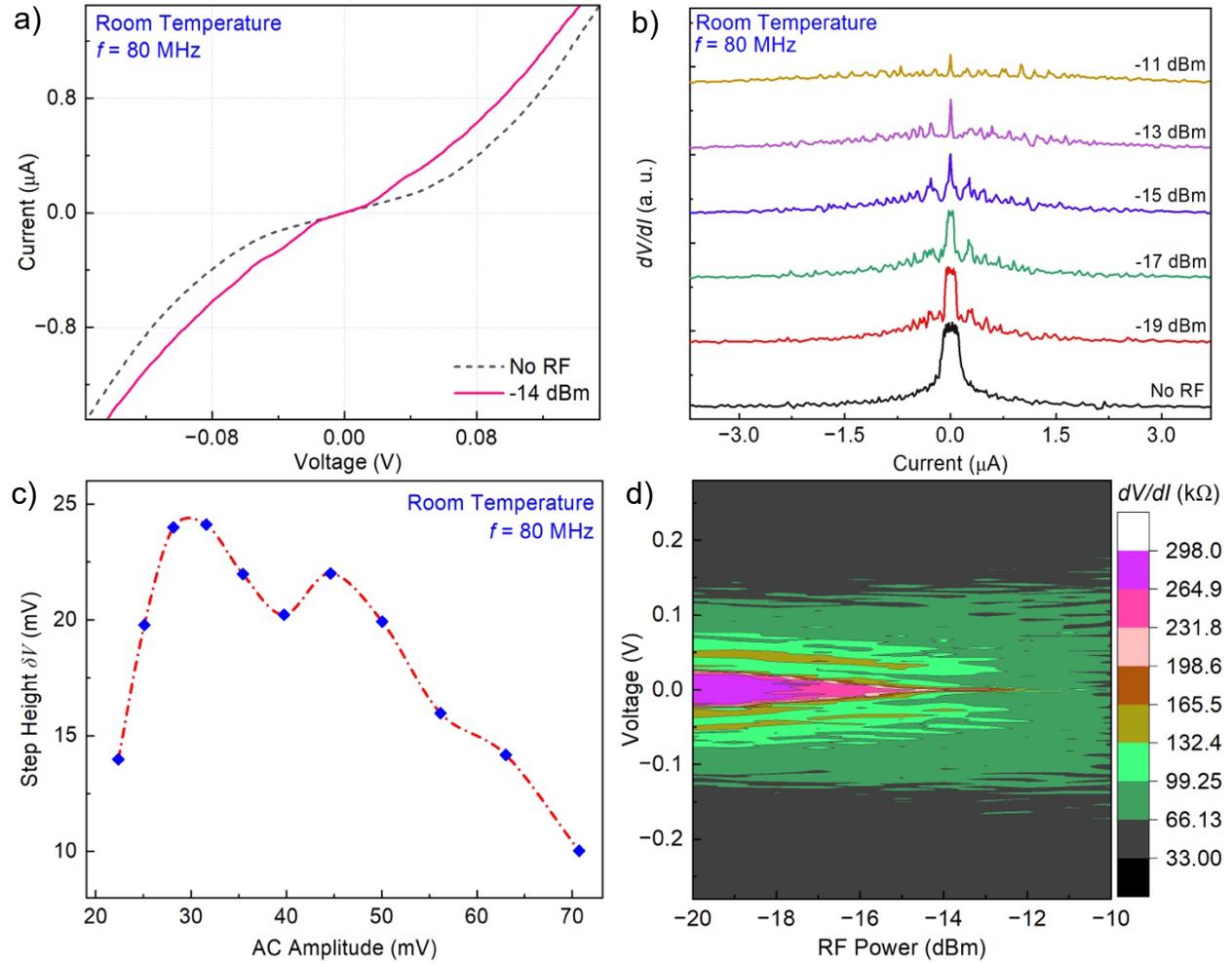

**Figure 3:** (a) The RT *I-V* characteristics without RF (dashed-black curve) and with RF (red curve) radiation of frequency *f* = 80 MHz and power $P_{RF}$ = -14 dBm. The ripples (red curve) are the Shapiro steps. (b) The differential *I-V* characteristics along the RF power range from $P_{RF}$ = -19 dBm to -11 dBm. The black bottom curve depicts the *dV/dI* signal when no RF signal is present. As the AC power increases, the harmonic peaks appear in *dV/dI* characteristics as a result of the CDW mode-locking with an external RF source. (c) The step height of the first harmonic plotted as a function of the RF amplitude at RT. (d) The color map shows the evolution of the peaks in differential characteristics as a function of the RF power and bias voltage.

In our experiments, the RF signal is mixed with the DC bias *via* a three-port Bias-Tee design. The output signal contains the AC component superimposed on the DC bias which sweeps through the selected range. Figure 3 (a) shows the *I-V* characteristic of the NbS$_3$ channel without RF input (dashed-black curve) and with the RF input of *f* = 80 MHz and power $P_{RF}$ = -14 dBm (red curve) at RT. One can see the Shapiro steps as a series of ripples in the *I-V* curve. The steps are observed even more clearly in the differential, *dV/dI*, characteristics presented in Figure 3 (b). The Shapiro steps occur due to the synchronization of CDW's fundamental frequency with the external RF source frequency, which results in the elimination of the condensate electrons from the current transport in the mode-locking state. As the RF power increases, the central peak narrows down, and extra harmonics appear. For example, at $P_{RF}$ = -15 dBm, the symmetric first harmonic peak is observed at ~ ±34 mV. The step height, *δV,* for the first harmonic peak, can be calculated by integrating the area under the *dV/dI* interference peak vs the drain current, *I*. Figure 3 (c) shows *δV* as a function of the AC amplitude at RT. One can notice an increasing trend in the step height at a small AC drive amplitude and an oscillating behavior as the AC amplitude becomes larger. This behavior agrees with previous reports of synchronization in CDW materials with quasi-1D crystal structures [18, 28]. For a better understanding of the evolution of the peaks with the RF power and bias voltage, the color map of the *dV/dI* values is presented in Figure 3 (d). The purple color indicates the central peak, while the two symmetrical dark yellow regions correspond to the first harmonic. As the RF power in the channel increases, *dV/dI* becomes smaller and the step height reduces. Our synchronization experiments provide additional proof that CDW quantum condensate phase is present in our samples, and can be controlled with the temperature, the source–drain bias, and the AC input.

We have experimented with the modulation of electrical current with the top gate in the *h*-BN/NbS$_3$ quasi-2D/1D heterostructure devices in the range of temperatures near RT. Figure 4 (a) shows the *I-V* characteristics for a representative device with a channel length of *L* = 4 µm at *T* = 220 K. In this measurement, the source-drain voltage is swept, while a fixed gate bias, $V_G$, is applied. The sweep rate is ~180 mV/s. In the linear *I-V* region, where the electrical current is comprised of the single particle transport, no modulation is observed with the source-drain voltage sweep. The modulation appears at the bias above the threshold voltage $V_T$, in the nonlinear region of the *I-V* characteristics. In this region, the current is dominated by the collective current of the sliding CDW quantum condensate. This suggests that, in this regime, we are modulating with the gate bias, the quantum condensate rather than the current of individual electrons. Figures 4 (b) and 4 (c) present a close view of the current for the positive and negative drain-source voltages, respectively. One can see that the gate bias effect on the sliding quantum condensate is similar and reversible for positive and negative source-drain biases. In Figure 4 (d), the reversibility for the gate bias is further illustrated with the current dependence on the gate bias in the forward and reverse gate voltage sweeping.

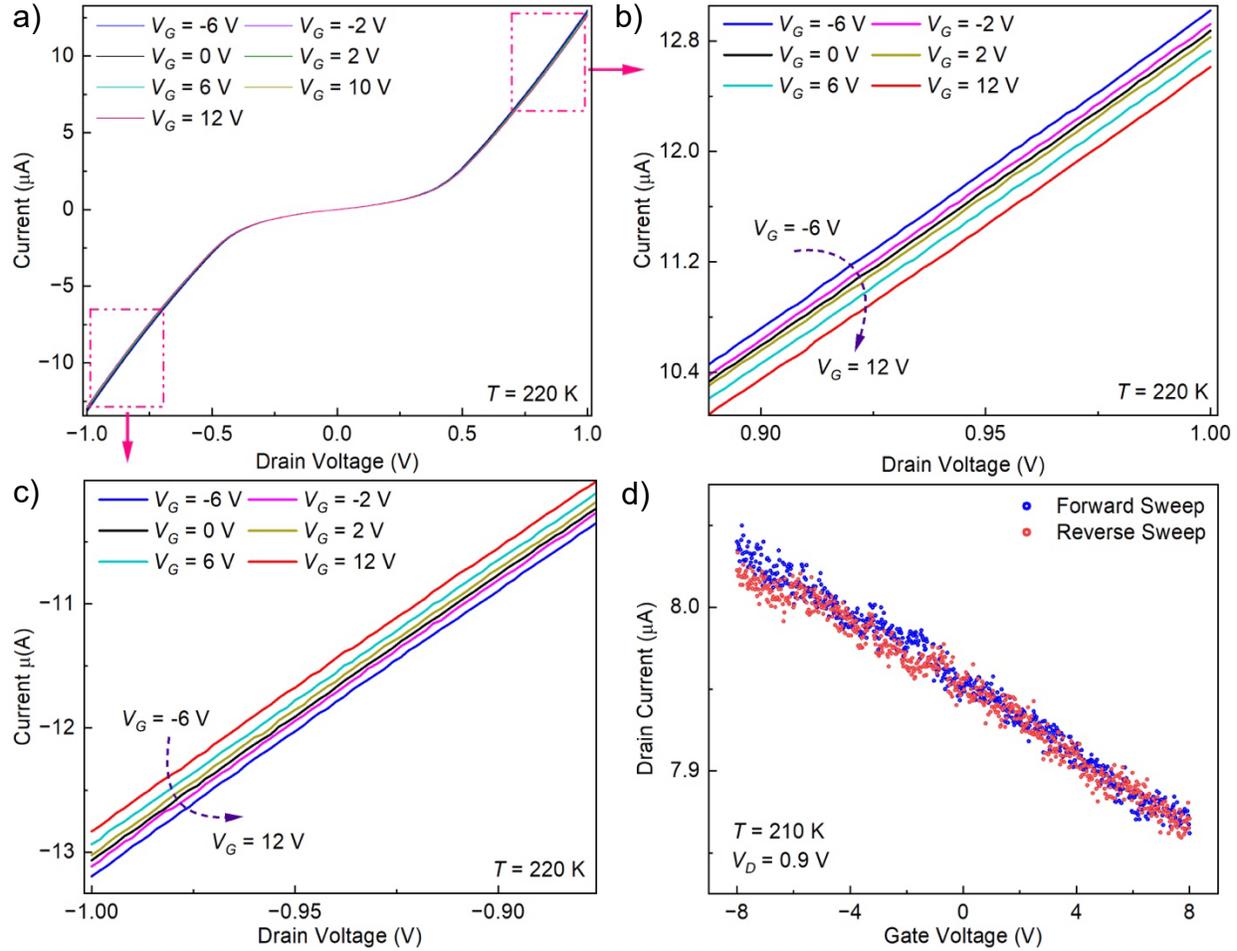

**Figure 4**: (a) The *I-V* characteristics of the *h*-BN/NbS$_3$ quasi-2D/1D heterostructure device under the fixed gate bias ranging from $V_G$ = -6 V to 12 V at $T$ = 220 K. The enlarged I-Vs near $V_D \approx$ 1 V $V_D \approx$ -1 V are shown in panels (b) and (c), respectively. (d) The drain current as a function of the forward and reverse gate voltage sweeps at the fixed source-drain bias of $V_D$ = 0.9 V. Note that the gate modulation of the sliding CDW quantum condensate is reversible.

We now examine closer the influence of the gate electric field on the *I-V* characteristics below and above the threshold field $E_T$. In Figure 5 (a), we show the linear *I-V* region (red-shaded area) with the current of single electrons, the intermediate near-threshold region (green-shaded area), and the non-linear *I-V* region (yellow-shaded area) with the collective current of the sliding quantum condensate. In the experiments with the pinned condensate (red-shaded region in Figure 5 (a)), the source-drain bias was fixed at a small value, in the range from $V_D$ = 10 mV to 45 mV, and the gate bias was swept across the channel with the rate of ~372 mV/s. The results are shown in Figure 5 (b). The current reveals only small-amplitude fluctuations, on the scale of ~ 1 nA, reminiscent of switching in the two-level systems. Supplemental Figure S4 shows that the leakage current in our devices is negligible and does not affect the measurements while Supplemental Figure S5 provides additional evidence for the switching behavior in the measurements with fixed gate bias. In the intermediate region, between $V_D$ = 50 mV and $V_D$ = 100 mV, just before and after the threshold field, one can see the evolution of the current response (see Figure 5 (c)). At a larger source-drain bias, $V_D \geq 0.3$ V, in the collective current regime of the sliding CDW condensate, we observe a clear gate field effect. Note that more positive bias resulting in a smaller current (see Figure 5 (d)). The current change due to the gate bias, in the devices with micrometer-long channels, is on the order of ~0.1 µA which is two orders of magnitude larger than the current variation in the single-particle regime.

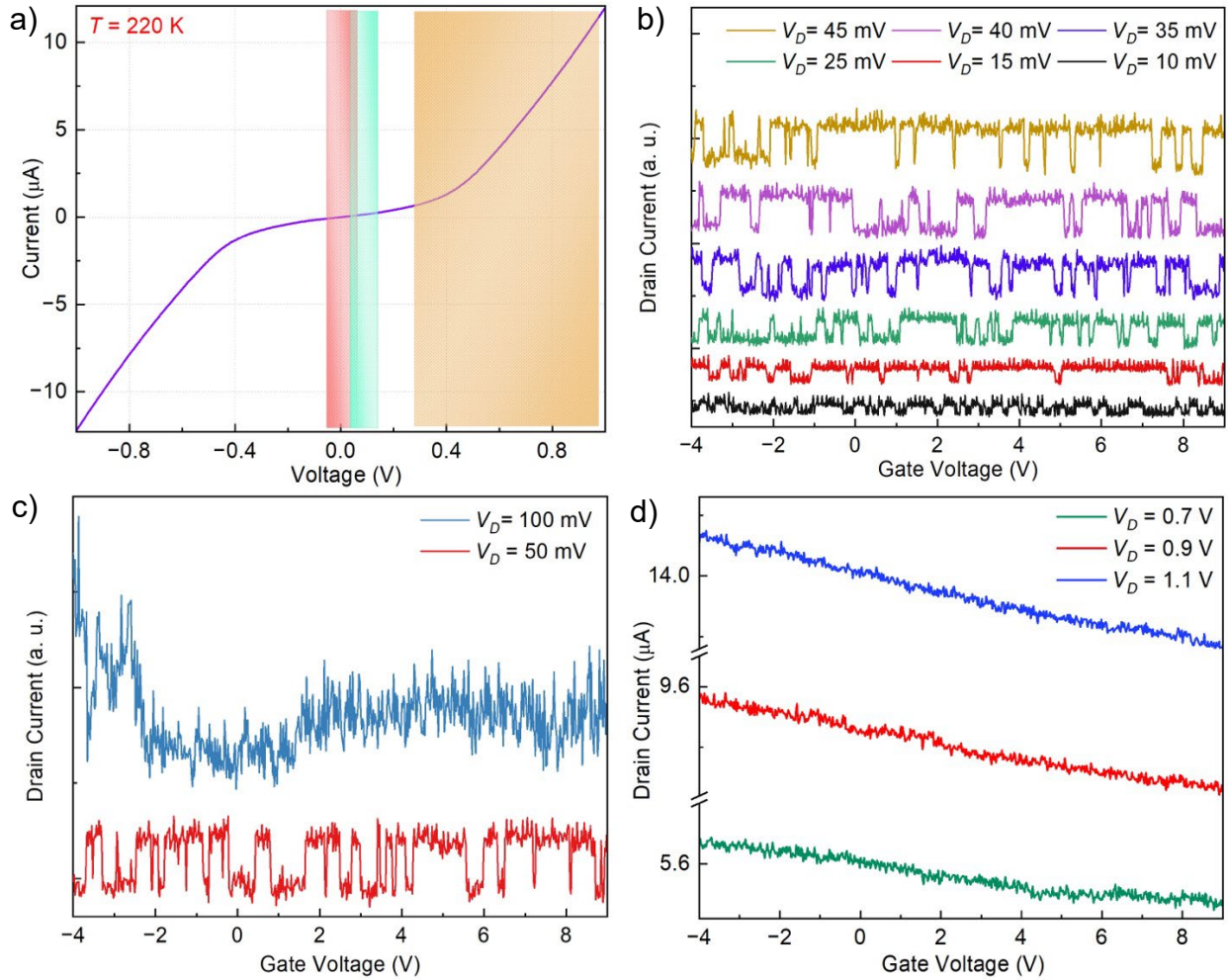

**Figure 5**: (a) *I-V* characteristics of *h*-BN/NbS$_3$ quasi-2D/1D heterostructure devices with the regions of the single electron current (red shade), intermediate near-threshold (green shade), and collective current (yellow shade) indicated. (b) Current as a function of the gate sweep voltage for the fixed small source-drain bias in the range from $V_D$ = 10 mV to 45 mV, corresponding to the single electron current regime. (c) Current as a function of the gate sweep voltage for the fixed source-drain bias in the near-threshold region. Note the disappearance of the switching type behavior at $V_D$ = 100 mV. (d) Current as a function of the gate sweep voltage for the fixed source-drain bias in the collective current region, characterized by the sliding CDW quantum condensate.

The electrical-field modulation of the sliding CDW quantum condensate that we achieved in *h*-BN/NbS$_3$ heterostructure devices is small but intriguing, and it has important implications. We have observed that the electric-field modulation is enhanced with the decreasing cross-section of the device channel. However, definitive conclusions require an additional systematic study, which can be possible with the improved technology of nanofabrication of test structures. Understanding the physical mechanism of the current modulation with the electrical gate in NbS$_3$ channels can potentially help enhance the effect in devices of different designs and material systems [23, 24]. We considered the possibility that the electrical field, created by the gate, changes the threshold field for the onset of the CDW sliding. However, we determined that in our top-gated short-channel devices, *i.e. L*~ 1 μm – 4 μm, for the $V_G$ change from -6 V to 12 V, the variation in $E_T$ was less than 30 mV. It is unlikely that this variation was behind the gate-field effect observed in our devices. Since the collective current reduces with more positive $V_G$, one can assume that the condensate attraction to the surface, or the transverse variation in the charge density, results in a decrease in its mobility. Typically, interfaces have a higher density of defects that can provide extra resistance to the CDW sliding. The small amplitude fluctuations observed in the current at low source-drain biases can be attributed to the variations in single particle conduction owing to the changing background potential experienced by individual electrons due to the presence of the pinned or creeping CDW. A possible effect on the current of the CDW creep, *i.e.* slow motion of CDWs with stops, or motion of segments, before its coherent sliding, was discussed in the context of other material systems [20-22]. Finally, one can mention that even though the electric-field modulation is small one should not exclude its potential for electronic applications – the CDW devices operate on different principles than FETs and may not need high on-off ratios achieved with the gate bias. The gate modulation observed near and at RT can be particularly beneficial for applications.

In conclusion, we reported the field-effect modulation of the sliding CDW quantum condensate in top-gated *h*-BN/NbS$_3$ quasi-2D/1D heterostructure devices near RT. The collective current reduces with more positive gate bias suggesting that the condensate attraction to the surface results in a decrease in its mobility. The knowledge of the electric-field effect on CDWs in quasi-1D NbS$_3$ nanowires in the sliding regime is useful for potential applications of such quantum materials in future electronics.

**Acknowledgments**

A.A.B. also acknowledges the Vannevar Bush Faculty Fellowship (VBFF) from the Office of Secretary of Defense (OSD), under the Office of Naval Research (ONR) contract N00014-21-1-2947 on One-Dimensional Quantum Materials. The work at UGA was supported by a subcontract from the VBFF ONR contract N00014-21-1-2947. The authors thank Dr. Sergey Rumyantsev, Dr. Fariborz Kargar, and Dr. Mykhaylo Balinskyy for their valuable discussions.

**Author Contributions**

A.A.B. conceived the idea, coordinated the project, contributed to experimental data analysis, and led the manuscript preparation; M.T. fabricated the devices, measured *I-V* characteristics, analyzed the experimental data, and wrote the initial draft of the manuscript; N.S. synthesized the material and conducted material characterization; T.T.S. supervised material synthesis and contributed to data analysis. All authors contributed to the manuscript preparation.

**Supplemental Information**

The supplemental information with additional microscopy data and *I-V* characteristics is available at the journal website free of charge.

**Data Availability Statement**

The data that support the findings of this study are available from the corresponding author upon reasonable request.